# The effect of variable fibre diameters in unidirectional fibre-reinforced bundles on stress redistributions around fibre breaks

M. Jafarypouria*, S.V. Lomov, S.G. Abaimov

*Skolkovo Institute of Science and Technology, Moscow, Russia*

*Corresponding author: Milad.Jafarypouria@skoltech.ru

**Abstract**

Finite element modelling is conducted to simulate the stress redistribution around a broken fibre (BF) in a bundle with experimentally measured fibre diameter distributions (FDD), followed by a parametric study of the influence of the FDD coefficient of variation on the stress concentration factor (SCF) and ineffective length (IL). Two variants of the SCF definition are considered: based on average and maximum stress in the fibre cross-section. Results demonstrate that bigger fibre diameters show higher SCF and clustering of such fibres increases SCF in nearest neighbour fibres (NNFs). Critically, maximum stress-based SCF ($maxSCF_{max}$) significantly exceeds average stress-based SCF ($maxSCF_{avg}$), with differences about 40–75% in NNFs for FDD bundles compared to FCD bundles. This emphasises the necessity of prioritizing maximum stress criteria over conventional average stress models in failure predictions. The findings challenge benchmark models that rely on averaged SCF values, offering critical insights for improving accuracy in predicting fibre break propagation and composite strength.

## 1. Introduction

Modelling of fibre breaks during the longitudinal tensile failure of unidirectional (UD) composites is a rapidly emerging area of research, currently the focus of several benchmark studies [1–3]. In the state-of-the-art models, the primary mechanism behind fibre break development involves the redistribution of load resulting from the tensile failure of individual fibres. This stress redistribution is characterized by SCF in neighbour intact fibres and IL of a broken fibre. As a fibre breaks, it can no longer support the load near the failure point, causing stress to be redistributed to neighbouring fibres. Unlike fibre-bundle models without a matrix, where a broken fibre is excluded from further analysis, the impregnated fibre bundle model (IFBM) allows for the failed fibre to still carry load, but only at distances along the fibre that exceed a defined ineffective length (IL). Close to the break point, this redistributed stress enhances the load on adjacent fibres, with the relative stress increase compared to the nominal stress termed the stress concentration factor (SCF). In impregnated fibre bundles, failure is characterized by spatially distributed and weakly correlated fibre breaks [4].

The fibre break model uses the stress redistribution around a single fibre break for a random packing of unidirectional fibres with a given fibre volume fraction (VF). The corresponding finite element (FE) simulations on a representative volume element (RVE) of the bundle are run beforehand; their output are the stress distribution profiles in the broken fibre and in the intact fibres surrounding it, depending on the distance of the fibre to the broken fibre. These stress redistributions are expressed using trend surface equations.

By the nature of this modelling a number of assumptions is used: (1) the FE geometry (or several random realisations of the geometry) represents in statistical sense the random placement of the fibres in the bundle; (2) the size of the model and the used boundary conditions lead to stress redistribution, which can be used to model a part of much larger bundle; (3) the trend surface equations for SCF are used in a deterministic way, neglecting deviation from the trends because of particular details of the fibre random placement.

*List of symbols*

| | | | |
|---|---|---|---|
| FDD | Fibre diameter distribution | max | maximum |
| FCD | Fibre constant diameter | $E$ | Young's modulus |
| SCF | Stress concentration factor | $\nu$ | Poisson's ratio |
| IL | Ineffective length | $\sigma_y$ | Yield stress |
| IFBM | Impregnated fibre bundle model | $L$ | RVE length |
| $\sigma$ | Experimentally measured standard deviation of the fibre diameter distribution | $\emptyset$ | RVE diameter |
| VF | Fibre volume fraction | $N_F$ | Number of fibres |
| $maxSCF$ | Maximum SCF | $N_E$ | Total number of elements in RVE |
| $D_{mean}$ | Mean fibre diameter | $\sigma_{z,avg}$ | average fibre stress |
| $D_i$ | Minimum fibre diameter | RFCD | Regression in the FCD case |
| $D_o$ | Maximum fibre diameter | $SCF_r$ | Regression of the reference SCF curve |
| CV | Coefficient of variation | $d/R$ | Relative distance to the broken fibre, with $R$ being fibre radius |
| RVE | Representative volume element | $\bar{\mu}$ | Mean |
| $D$ | Fibre diameter | $p_{value}$ | Significant statistical difference |
| Avg | Average | BF | Broken fibre |
| Std | Standard deviation | NNFs | Nearest neighbour fibres |
| min | minimum | | |

The present work deals with this FE modelling, in the paradigm, established in the research of the past decade, by efforts of different research groups, cited above [1–3]. The modelling, described here, is based on the assumptions, listed above, and uses particular choices for the models creation, which are established in [1–3]. These choices include most importantly size of the RVE; boundary conditions for the RVE; stress transfer conditions near the fibre break. The reader is referred to [5] for detailed discussion of these subjects. The step beyond the state-of-the-art is in considering a bundle of fibres with different diameters.

The stress redistribution around broken fibres in fibre break models has been addressed by considering factors such as non-linear matrix behaviour [6–9], dynamic stress concentrations arising from the abrupt brittle failure of fibres [6,10], random fibre packing [6,7,10–13], and fibre misorientation [14]. However, to our knowledge, the influence of fibre diameter distribution (FDD) on the SCF and IL has not yet been explored. While previous studies [10,11,15–17] have incorporated different fibre diameters for various fibre types in hybrid composites, they have not accounted for variability in diameter within a single fibre type.

The diameters of fibres can be measured using a variety of tools and techniques. Mesquita and et al. [18,19] utilized automated single fibre tensile testing to generate extensive datasets of carbon and glass fibre properties, measuring fibre diameter with a laser diffraction system. These datasets are used as the foundation for FDD characterization in this study.

Composite fibre spatial distribution is affected by production conditions like pressure, matrix flow, and mould constraints. Even under ideal manufacturing conditions, fibres are irregularly placed, leading to clustering and resin-rich areas. Statistical descriptors can quantify this irregularity [20–22], including fibre position, diameter distribution, and alignment. The most relevant useful parameter is the fibre diameter [23]. We apply the Melro – Catalanotti adapted algorithm [24,25] for fibres of constant diameter (FCD) and fibres following a diameter distribution (FDD) based on validation [21]. This algorithm is used to create random fibre configurations and analyse the differences between the FCD and FDD cases for carbon bundles using distributions obtained from [18]. Fibre orientation's effect is recognized as an influential parameter [26] and is investigated by the present authors for the FCD case [14], but not included in the current study.

Zangenberg et al. [21] carried out a full statistical analysis for fibre geometry descriptors based on (i) number of neighbours, (ii) nearest neighbour distance, (iii) contact points per fibre, and (iv) local fibre volume fraction. Through a parametric variation of the fibre radii distribution (mean, standard deviation, and skewness), they found that, when taking into account the entire range of inter-fibre distances, there is very little difference between the descriptors for different fibre radii distributions. Nonetheless, the microscopic stress distribution is extremely sensitive to the inter-fibre distances when fibres are close one to another [27]. The SCF may "feel" the differences in the nearest fibre positions, which are smoothed away when the edge bins of the distribution are evaluated. Consequently, when the various fibre diameters are taken into account, one may anticipate that SCF will still show a difference despite the findings in [21].

The baseline models [1–3] use the average SCF taken over the cross-section as the maximum stress in the neighbour intact fibre. An alternative approach is to use the maximum stress in a cross-section as a criterion for the fibre break [28–30], especially because the strong stress localisation is supported by experimental evidence [28]. The axial stress gradient in the radial direction is very strong and the stress decays very quickly in the radial direction from the broken fibre. This is expected to be even more important in our study of fibres with different diameters. Therefore, in the present study, we aim to calculate the SCF values in both formats (average and maximum over NNF's cross-section) and compare them.

The objective of the present study is to investigate the influence of FDD, in comparison to FCD, on the stress redistributions (SCF and IL), based on finite element (FE) analysis of the stress state of a bundle around a broken fibre. The investigation starts with the case of experimentally observed FDDs for carbon fibres and is followed by parametric analysis of the SCF and IL change for different FDD widths. The comparison of FCD vs FDD bundles, for SCF and IL, is conducted for: (I) elastic and inelastic regime of deformation, (II) average SCF and SCF based on maximum stress.

## 2. Fibre diameter distribution: Experimental data and numerical generation

### 2.1. FDD data for carbon T700s fibres

The current study uses statistical data for the diameter distribution of carbon T700s fibres reported in [18,19] and represented in Table 1. It is assumed that the distribution of diameters is Gaussian. A histogram of fibre diameters for carbon T700s fibres is displayed in Fig. 1 [18,19].

In order to choose an appropriate probability distribution type, let us suppose that the diameter values follow the experimental distribution with the measured standard deviation $\sigma$ reported in [18] and represented in Table 1. Fig. 1 illustrates an observed variation in the fibre diameter of carbon fibres and fits for typical probability distributions (gamma, skew-normal, log-normal, and normal).

Fig. 1 shows a small difference between the fits of various probability distributions. In contrast to the normal distribution, three probabilities—skew-normal, log-normal, and gamma—display a left-skewness and are indistinguishable from one another. As the comparison of the fitted distributions with the experimental histograms depicts, the data do not clearly favour one distribution over another. Thus, the normal distribution is used in the current paper's calculations.

**Table. 1.** Parameters of carbon T700s fibres used to generate FDD diagrams [18].

| Definition | T700s fibre |
|---|---|
| Mean fibre diameter ($D_{mean}$, μm) | 6.76 |
| Minimum fibre diameter ($D_i$, μm) | 6 |
| Maximum fibre diameter ($D_o$ μm) | 7.8 |
| Coefficient of variation (CV) | 4.6% |

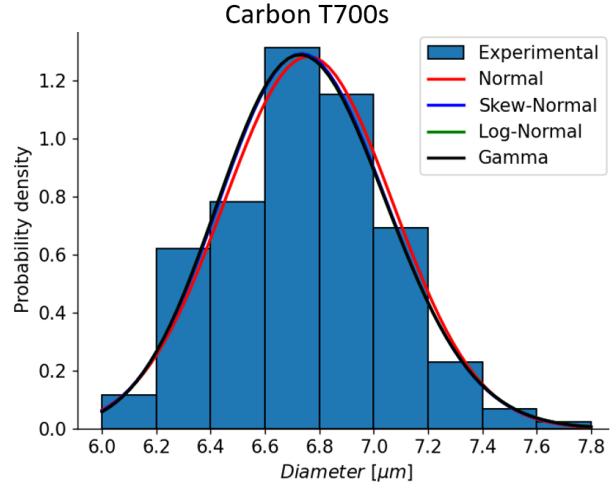

**Fig. 1.** Fibre diameter distributions, following [18,19], and fits to common probability distributions (the skew-normal, log-normal, and gamma distributions are indistinguishable).

## 2.2. Random fibre placement

In order to generate a fibre spatial distribution, a modified version of the Melro-Catalanotti algorithm [24, 25] is used. During subsequent algorithm steps, fibres with different diameters are implanted into the RVE in accordance with the assumed fibre diameter distribution.

In the state-of-the-art models, the RVE needs to be large enough in the fibre longitudinal direction to guarantee that the ineffective length will be well captured. RVEs with lengths ranging from 7.5×$D$ to 20×$D$ were used in a number of studies [7,12,15], where $D$ is the fibre diameter. In the present study, the size of the RVE was selected as 12×$D$ and 20×$D$, where $D$ is the mean fibre diameter (Table 1), for transversal and longitudinal directions, respectively. The authors of [12,15] have shown that using the same RVE dimensions guarantees the representativity of SCF calculations for the range of fibre volume fractions under consideration.

Fibres were arranged in a square area before being cut into a cylindrical bundle, potentially resulting in a different fibre volume fraction (VF) compared to the target VF in the square representative volume element (RVE). The real volume of the fibres that are "cut" by the RVE boundary should be taken into consideration when calculating the fibre volume in the cylindrical RVE. The "fibre volume/RVE volume" ratio in Abaqus was used to compute the actual VF in our models. The real VFs in the cylindrical RVE with carbon bundles for the FCD and FDD models are displayed in Table 2. The FCD and FDD models' minimum (Min), maximum (Max), average (Avg), and standard deviation (Std) values are displayed in this table.

FCD and FDD IFBMs were created, where RVEs included three nominal fibre VFs, namely 30%, 50%, and 65% (see Fig. 2). Five realisations for FCD and FDD bundles were built for statistical analysis.

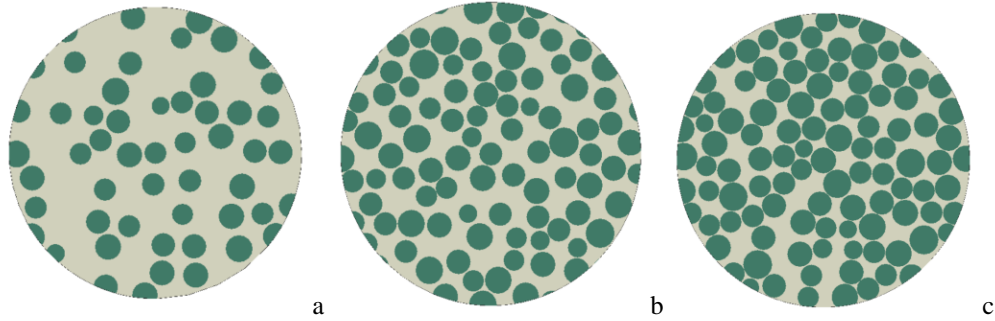

**Fig.2.** Examples of fibre placements generated by the presented algorithm: a) VF30%, b) VF50%, and c) VF65%.

**Table. 2.** Actual fibre volume fractions in the cylindrical bundle cut from square one; data were extracted based on "fibre volume/RVE volume" ratio in Abaqus.

|        |               | FCD models |          |      |      | FDD models |          |      |      |
|--------|---------------|-----------|----------|------|------|-----------|----------|------|------|
| Fibre  | Nominal VF    | Min VF %  | Max VF % | Avg  | Std  | Min VF %  | Max VF % | Avg  | Std  |
|        | 30%           | 29.4      | 30.8     | 29.9 | 0.66 | 30.0      | 33.4     | 31.6 | 1.20 |
| Carbon | 50%           | 48.2      | 51.1     | 49.9 | 1.09 | 48.0      | 51.1     | 49.7 | 0.96 |
|        | 65%           | 63.6      | 66.1     | 64.9 | 1.02 | 63.6      | 66.8     | 64.7 | 1.08 |

## 3. Materials, finite-element models, and plan of numerical experiments

Table 3 presents the material properties of carbon T700s fibres [12] and the epoxy matrix [31]. The fibres are modelled as perfectly elastic, while the epoxy matrix exhibits elasto-plastic behaviour with linear hardening.

The RVE used to analyse the stress redistribution around a single fibre break in a randomly packed arrangement of parallel fibres is depicted in Fig. 3a,b. Fig. 3c illustrates the boundary conditions, and Fig. 3d provides an example of the longitudinal stress concentration in intact carbon fibres adjacent to the fibre break, with a refined mesh applied to the fibres. The broken fibre (BF) and its nearest neighbour fibres (NNF) are highlighted. The finite element (FE) simulation is based on the model established by Swolfs et al. [12].

The whole black plane (Fig. 3c) of each IFBM experiences a displacement of $0.001 \times L$ and $0.02 \times L$ under uniaxial tension ($L$ is the RVE length). The RVE's lateral surface is configured to be traction-free. In order to simulate the break, z-symmetry boundary condition is applied to RVE at the plane of the fibre break (yellow plane), with the exception of the central fibre's cross-sectional area and its perimeter (which are coloured gray). This yields a realistic SCF profile when compared to the Raman spectroscopy data [32], as seen in [13]. See Table 4 for further information regarding the FCD and FDD FE models.

Based on a mesh sensitivity study, the broken fibre, the surrounding matrix, and the section of neighbouring fibres facing the broken fibre are allocated a denser mesh (with the smallest element size of 60 *nm*×200 *nm*×360 *nm*). The mesh is progressively coarsened out axially and radially away from the broken fibre and its plane, taking into account the computational efficiency (see Fig. 3). Mesh verification supporting this assumption is presented in Supplementary Materials, Section SUPP 1.

**Table. 3.** Material properties of carbon T700s fibre and epoxy matrix.

| Carbon T700s fibre (Transversely isotropic) [12] | | | | | | | | |
|---|---|---|---|---|---|---|---|---|
| $E_{11}$ (GPa) | $E_{22}$ (GPa) | $E_{33}$ (GPa) | $\nu_{12}$ | $\nu_{13}$ | $\nu_{23}$ | $G_{12}$ (GPa) | $G_{13}$ (GPa) | $G_{23}$ (GPa) |
| 230 | 15 | 15 | 0.25 | 0.25 | 0.25 | 13.7 | 13.7 | 6 |
| Epoxy Matrix (Elasto-plastic with linear hardening [31]) | | | | | | | | |
| Initial stiffness ($E_0$) | | | | | | 2.95 GPa | | |
| Hardening stiffness ($E_1$) | | | | | | 0.82 GPa | | |
| Poisson's ratio ($\nu$) | | | | | | 0.4 | | |
| Yield stress ($\sigma_y$) | | | | | | 66 MPa | | |

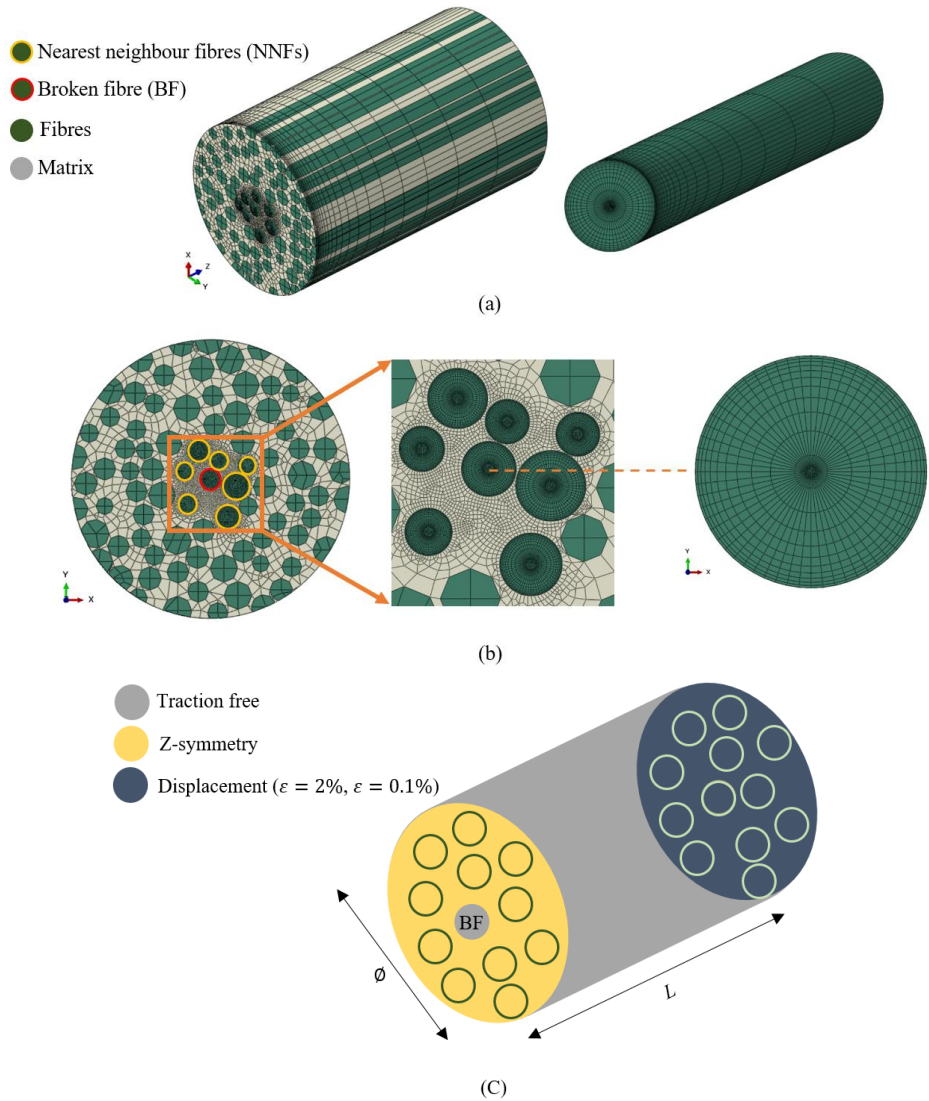

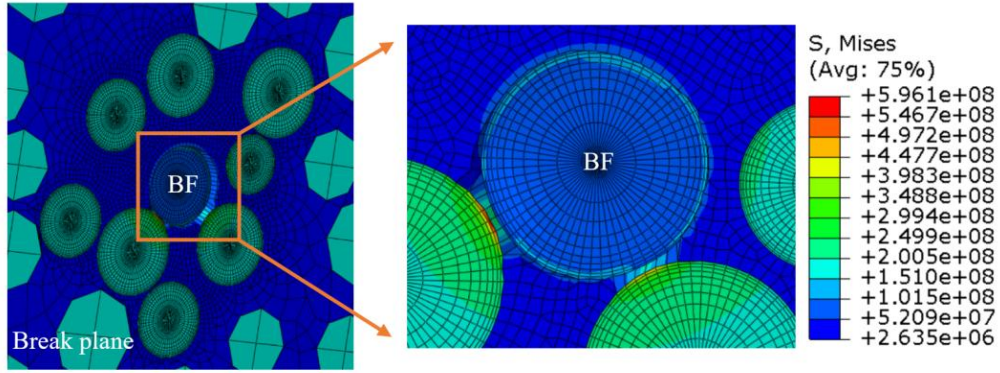

(d)

**Fig.3.** FE models: a) mesh density along the fibre axis, b) mesh density on the fracture plane, c) boundary conditions and model dimensions, and d) failing fibre and its neighbours along with the longitudinal stress field at the plane of the fibre break. Applied strain 0.1%, VF = 50%.

**Table. 4.** Parameters of the finite element simulations of FCD and FDD models.

| Parameters | FCD bundles | FDD bundles |
| --- | --- | --- |
| Carbon fibre diameter ($D$) | 7 $\mu m$ | $D_{mean} = 6.76$ μm |
| $L$ (RVE length) | 20×$D$ | 20× $D_{mean}$ |
| $\emptyset$ (RVE diameter) | 12×$D$ | 12× $D_{mean}$ |
| $N_F$ (Number of fibres) | 45 fibres for $V_f = 30\%$<br>75 fibres for $V_f = 50\%$<br>100 fibres for $V_f = 65\%$ | 60 fibres for $V_f = 30\%$<br>100 fibres for $V_f = 50\%$<br>130 fibres for $V_f = 65\%$ |
| $N_E$ (Total number of elements in RVE) | 450,000 < $N_E$ < 800,000 | 450,000 < $N_E$ < 800,000 |
| Type of elements | 85% - 90% first-order hexahedral elements<br>10% - 15% first-order wedge elements | 85% - 90% first-order hexahedral elements<br>10% - 15% first-order wedge elements |

From the stress field in 3D FE models, the IL of a broken fibre and the SCF values of the neighbouring intact fibres were determined. Rosen [33] states that twice the fibre length over which 90% of stress recovery takes place is the ineffective length (see Fig. 4a).

For average SCF, the definition by Swolfs [12] was used: the SCF is calculated as the relative increase in average fibre stress $\sigma_{z,avg}$ at $z^*$ divided by the average fibre stress $\sigma_{z,avg}$ far away from the failure plane:

$$SCF_{avg}(z = z^*) = \frac{\sigma_{z,avg}(z = z^*) - \sigma_{z,avg}(z = L)}{\sigma_{z,avg}(z = L)} \times 100 \tag{1}$$

where $\sigma_{z,avg}$ is defined as the longitudinal fibre stress of the elements in that plane averaged over the cross-section of the fibre and $z^*$ is the distance along the fibres from the break plane. According to Multiplier 100, all SCF results will be displayed as percentages. Fig. 4b shows how so-defined SCF changes as a function of distance along the length of the intact fibre, which is the closest neighbour of the broken fibre. The stress rapidly decays by moving away from the break plane once the maximum of the average SCF along the fibre ($maxSCF_{avg}$) is attained very close to the break plane [12].

In addition to the average SCF previously mentioned, we also take into account the maximum SCF, which is determined by the maximum stress at the cross-section of the fibre. In Eq. (1), $\sigma_{z,avg}$ is substituted with $\sigma_{z,max}$, which is the highest axial stress value in the fibre cross-section. Along the fibre, the maximum ($maxSCF_{max}$) is once more sought after. Fig. 4b, inset, illustrates the location of this peak and the stress distribution on the fibre cross-section.

Plastic zones can develop inside the polymer matrix when the applied load causes the matrix to deform more than its elastic limit. The greatest primary stress in the matrix surrounding the fibre break in the current model shown in Fig. 4 is 89 MPa, while its yield stress is 66 MPa. It signifies that the matrix has entered the plastic deformation regime and surpassed its yield stress. Since there has been plastic deformation of the matrix in this instance, matrix plasticity has emerged within the RVE.

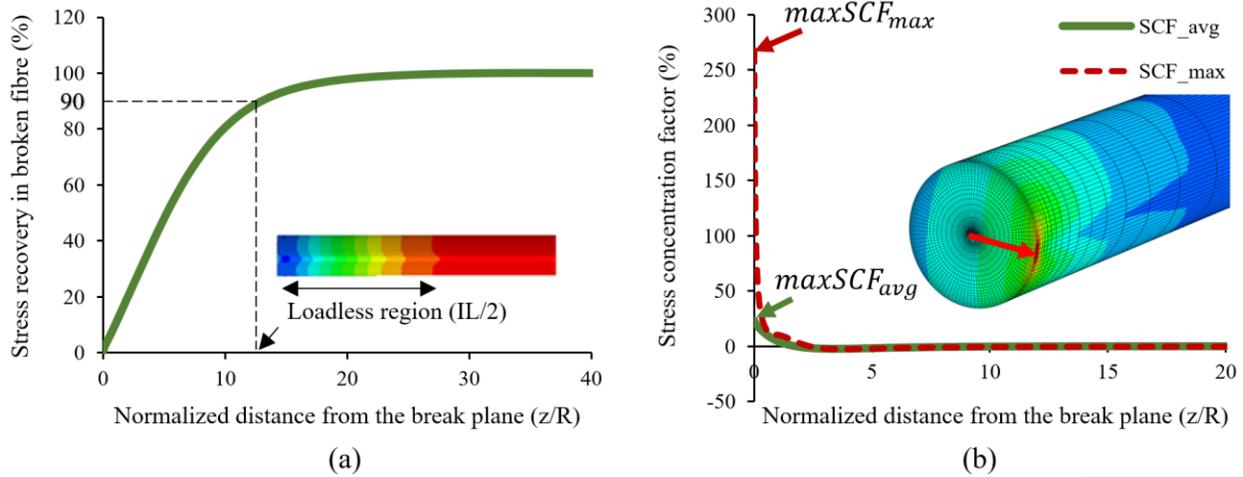

**Fig.4.** Stress redistribution: a) IL along the broken fibre, inset: axial stress on the surface of the broken fibre; b) $maxSCF_{avg}$ and $maxSCF_{max}$ along the nearest intact neighbour; inset: stress distribution on the cross-section of the intact fibre and along its surface. Applied strain 0.1%, VF50%.

To investigate the effect of the FDD and, specially, of the distribution standard deviation $\sigma$ on the stress redistribution around a single fibre break, several cases are compared: 1) FCD: $\sigma = 0$; 2) FDD with standard deviation of $\sigma/2$, $\sigma$, and $2\sigma$, where $\sigma$ is the experimentally measured standard deviation. The statistical comparison of the SCFs presented in SUPP 2 follows a one tail t-Test with an unequal variance analysis.

To evaluate the influence of the FDD on the stress redistribution around a single fibre break and judge the statistical confidence of the differences, the following approach was used. Let $SCF_r(d/R)$ be a regression of the reference SCF curve (baseline FCD with $\sigma = 0$) where $d/R$ is relative distance to the broken fibre, $R$ is fibre radius, and $SCF_k(d_k/R)$ with $k = 1...K$ is a set of $K$ calculated SCF values for fibres in a model of another type (FDD with $\sigma/2$, $\sigma$, or $2\sigma$). Then statistical analysis is conducted on the differences $\Delta_k = SCF_k(d_k/R) - SCF_r(d_k/R)$, analysing a hypothesis $\Delta_k \neq 0$.

## 4. Results and discussion

This section presents the stress redistribution (SCF and IL) for three nominal fibre volume fractions (VF) of 30%, 50%, and 65% for impregnated fibre bundles in FCD and FDD models. The SCF is calculated based on average normal stress ($maxSCF_{avg}$) and peak normal stress ($maxSCF_{max}$) over the cross-section.

## 4.1. Average normal stress over the fibre cross section

### 4.1.1. FCD models

Five random realisations were generated for carbon epoxy bundle. Tables 3 and 4 present the FE parameters and material properties that are used in the FCD models.

Fig. 5 shows the variations in stress concentrations calculated with the applied average strain of 0.1%, i.e. in the elastic regime of deformation, and at 2%, which determine the full effects of the non-linearity of the matrix behaviour. It depicts the distribution of maximum average SCF $maxSCF_{avg}$ over the fibres of one of random realisations. Only the small number of fibres, the closest neighbours of the broken fibre (see an example in Fig. 3b), have values of maximum SCF more than 1%, but these fibres are the most susceptible to break under the overstress caused by the stress redistribution. The difference in the loading level is felt primarily by these fibres. The maximum SCF (the rightmost bins in Fig 5) is decreased for 2% loading in comparison with 0.1%. The detailed analysis in [13] shows that the transition to the inelastic regime strongly affects the fibre breaks development. In the rest of the calculations, the applied average strain level of 2% is used, as it was done in the fibre break benchmarks [1–3].

Fig. 6a displays the ineffective length at different VFs. An increase in fibre volume fraction leads to a reduction in the IL due to increased shear stresses in the matrix, which arise from the decreased distances between fibres at higher volume fractions.

Fig. 6b illustrates the stress concentration trends for five random realisations of the FCD bundles at different VFs in the intact fibres surrounding the broken fibre. The $maxSCF_{avg}$ is represented as a function of the relative distance (*d/R*) from the fibre break. It is observed that SCF decreases rapidly as *d/R* increases, a phenomenon known as the 'shielding effect' [12]. In this context, the first layer of nearest neighbours exhibits a significantly higher SCF compared to the second layer of nearby fibres. This occurs because the closest fibres effectively shield their second-nearest neighbours from stress concentration. Consequently, the SCF for these second-closest neighbours is considerably lower. Consequently, the SCF of the second-nearest neighbours is considerably lower. Additionally, SCF values are higher at lower VFs, again attributed to the shielding effect; a higher fibre volume fraction means more nearby fibres, leading to a more pronounced shielding effect.

The FCD calculation results align closely with those presented in [1,12,13], providing confidence in their application as a baseline for further analysis of the effect of fibre diameter distribution.

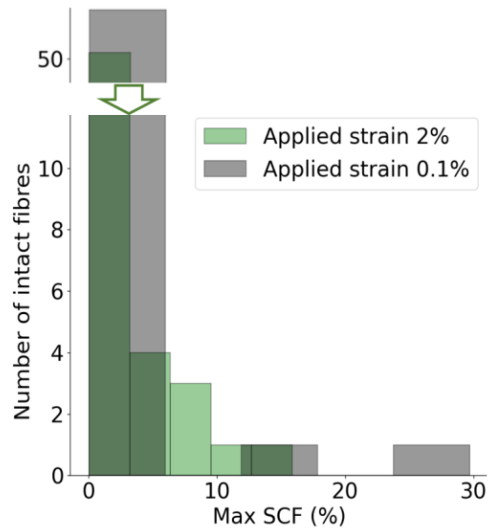

**Fig.5.** Maximum average SCF ($maxSCF_{avg}$) distribution in all intact fibres around fibre break in one of the random realisations of the FE model at VF50%, FCD.

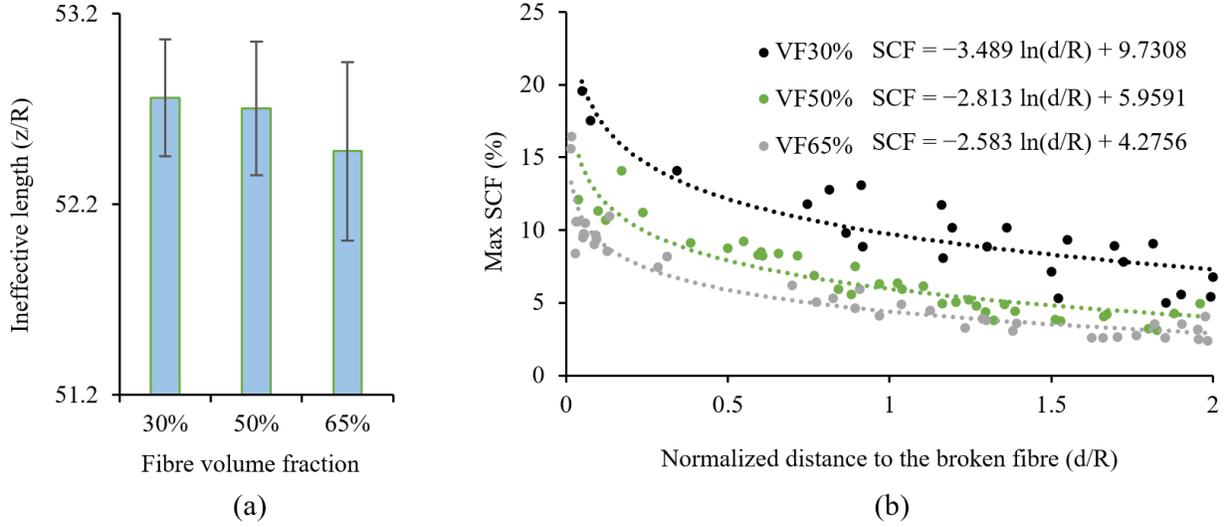

**Fig. 6.** Stress redistribution for FCD carbon T700s bundles at different VFs, five random fibre placement realisations for each VF: (a) IL, error bars show the standard deviation, (b) maximum average SCF ($maxSCF_{avg}$) in function of the normalised radial distance to the broken fibre, each data point represents the maximum SCF in one intact fibre in one of the realisations of the FE model. Applied strain 2%.

### 4.1.2. FCD vs FDD

A parametric analysis with three levels of the standard deviation ($\sigma/2$, $\sigma$, and $2\sigma$) was carried out to assess the effect of diameter variation on the stress redistribution surrounding a single fibre break, where σ is the standard deviation of the experimental data presented in [18]. As was covered in section 2, the parametric investigation was conducted using an assumed normal probability distribution. Five random realisations of the carbon bundles for each standard deviation were created using the normal distribution's parameters in accordance with the material characteristics and FE parameters listed in Tables 3 and 4.

Stress concentrations, Eq (1), for FCD and FDD bundles at different VFs in the neighbouring intact fibres surrounding a single fibre break are compared in Fig. 7. The maximum average SCF ($maxSCF_{avg}$) is shown as a function of the relative distance ($d/R$) from the fibre break.

Fig. 7a,c,e show SCF trends for FDD ($\sigma/2$, $\sigma$, and $2\sigma$) bundles in comparison with FCD ($\sigma = 0$) bundles. The peak of the $maxSCF_{avg}(\sigma)$ graph ($maxSCF_{avg}$ in NNFs) illustrates a notable decrease at VF65% of about –(20%-30%) for FDD bundles compared to FCD bundles. The $maxSCF_{avg}$ in NNFs of FDD bundles with $2\sigma$ shows an increase of 13% and 18% at VF30% and VF50%, respectively. A higher fibre volume fraction (VF65%) leads to more pronounced differences in the peak of the *maxSCF* because fibres are located closer to the broken fibre. Additionally, the peak SCF tends to decrease gradually as the volume fraction increases.

Fig. 7b,d,f exhibit the difference between the equivalent regression in the FCD case (RFCD) and the FDD bundle. Greater scattering of SCF data points for FDD bundles as opposed to FCD models results from the diversity in fiber diameter within FDD bundles, which causes significant changes in the normal stress experienced by neighboring fibers. This effect is most noticeable close to fiber break. The data points demonstrate a positive divergence from RFCD at volume fractions of VF30% and VF50% and a negative deviation at VF65%.

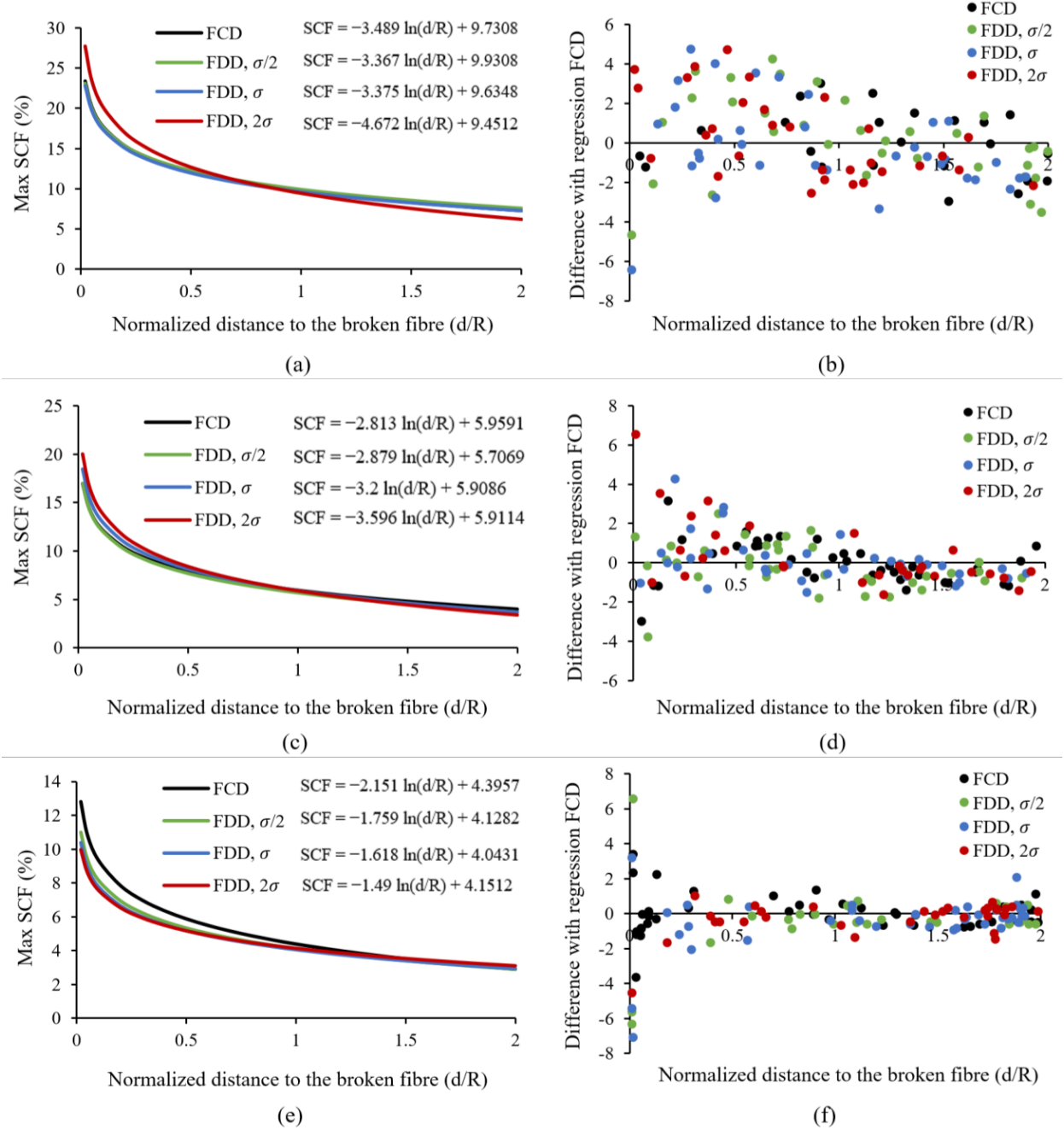

**Fig.7.** Stress concentrations for the FDD carbon T700s bundles compared with the FCD bundles for different VFs at applied strain 2%, five random fibre placement realisations for each VF: (a, c, e) SCF regressions in function of the normalised radial distance to the broken fibre and (b, d, f) maximum average SCF ($maxSCF_{avg}$) difference with regression in the FCD case ($FDD_{maxSCFs}$-RFCD), each data point represents the difference $FDD_{maxSCFs}$-RFCD in one intact fibre in one of the realisations of the FE model.

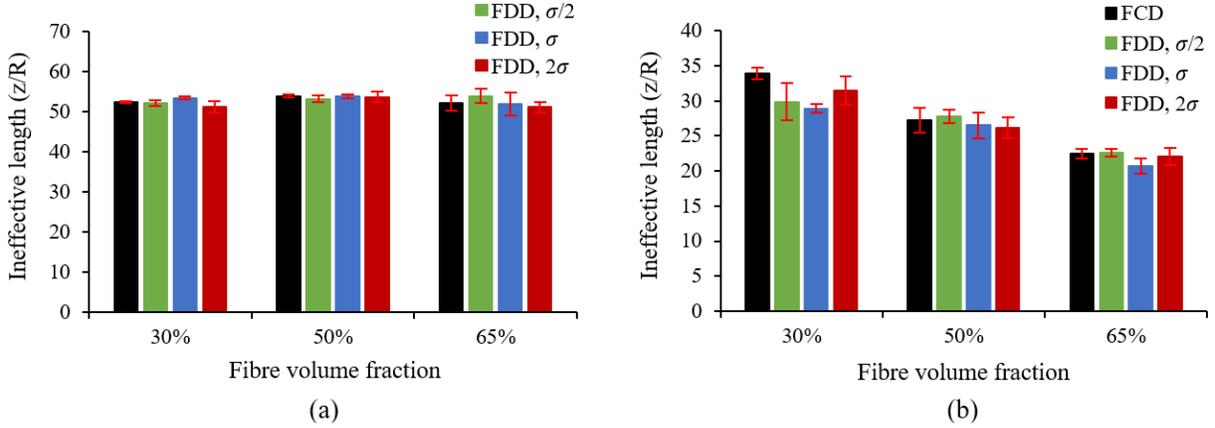

**Fig.8.** Ineffective length for the FDD bundles compared with the FCD bundles at different VFs: a) applied strain 2%, b) applied strain 0.1%. Five random fibre placement realisations for each VF, error bars show the standard deviation.

Increasing the ineffective length due to an applied strain of 2% (see Fig. 8) can lead to a decrease in the stress concentration factor. This occurs because a longer ineffective length facilitates a more gradual transfer of stress from the fibres to the matrix. Essentially, as the ineffective length increases, the stress can be dispersed across a larger area, decreasing the concentration of stress at any one location. Both localized failures and SCFs are significantly decreased by this distribution.

The IL of FDD and FCD bundles at various VFs is assessed in Fig. 8. At a strain of 2%, the ILs show little variation among various VFs and provide no discernible trend. However, the IL drops as the VF increases from 30% to 65% at a strain of 0.1%. At 2% strain, yielding close to the broken fiber reduces the matrix's capacity to transmit shear stress, which causes the broken fiber's stress recovery length to increase compared to the 0.1% strain condition. Furthermore, the shear stress encountered in the matrix between fibers varies significantly due to the variety in fiber diameters within the FDD bundles, resulting in distinct stress recovery behaviors for broken fibers in FDD bundles.

### 4.2. Maximum normal stress over the fibre cross section

#### 4.2.1. FCD vs FDD

In this section, the SCF is calculated based on the peak axial stress ($\sigma_{z,max}$) over the fibre's cross-section. The SCF at a particular *z*-coordinate along an intact fibre ($z^*$) with a total length of *L*, is the relative increase of the cross-sectional peak stress, $\sigma_{z,max}$, at $z = z^*$ with respect to the far-field stress (at $z = L$) ( $SCF_{max}(z = z^*) = \frac{\sigma_{z,max}(z=z^*) - \sigma_{z,max}(z=L)}{\sigma_{z,max}(z=L)} \times 100$). Then maximum is found along the fibre as $maxSCF_{max}$. Five random realisations of the carbon bundles for each VF were generated based on the material properties and FE parameters presented in Table 3 and Table 4. For the FDD bundles, the standard deviation $2\sigma$ was selected, as it demonstrated the maximum difference with FCD bundles in section 4.1.2.

Fig. 9a,c,e illustrate the $maxSCF_{max}$ trends for FCD and FDD bundles across VFs. At normalized radial distances $d/R < 0.5$, FDD bundles exhibit a 43–75% increase in peak $maxSCF_{max}$ compared to FCD, depending on VF. This heightened stress concentration arises because bigger-diameter fibres in FDD bundles bear proportionally higher loads due to their increased cross-section area, amplifying localized stresses near the break. In contrast, fibres with smaller diameter experience reduced stress concentrations, but their influence is overshadowed by the dominance of bigger fibres in proximity to the break. The rapid decay of $maxSCF_{max}$ with increasing $d/R$ (Fig. 9a,c,e) reflects the acute sensitivity of peak stresses to immediate fibre geometry variations, which diminish as distance from the break grows.

Fig. 9b,d,f show a clear positive deviation of $maxSCF_{max}$ in FDD bundles compared to FCD regressions (RFCD) at close relative distances ($d/R < 0.5$) to the fibre break. This variability highlights the necessity of considering random fibre diameter distributions in failure models. Moreover, there is a faster decay of FDD $maxSCF_{max}$ (compared to FCD) which is attributed to stress redistribution among fibres of differing diameters, which distributes localized stresses more effectively than uniform-diameter configurations.

The significant increase in $maxSCF_{max}$ observed in FDD bundles illustrates the limitations of average stress-based criteria (e.g., [1–3]). The maximum stress criterion identifies the fibre break initiation probability more effectively, particularly in composites with variable fibre diameters. These findings are consistent with experimental evidence of stress localization on fibre surfaces [28,29], validating the need to prioritize $maxSCF_{max}$ in a bundle strength model.

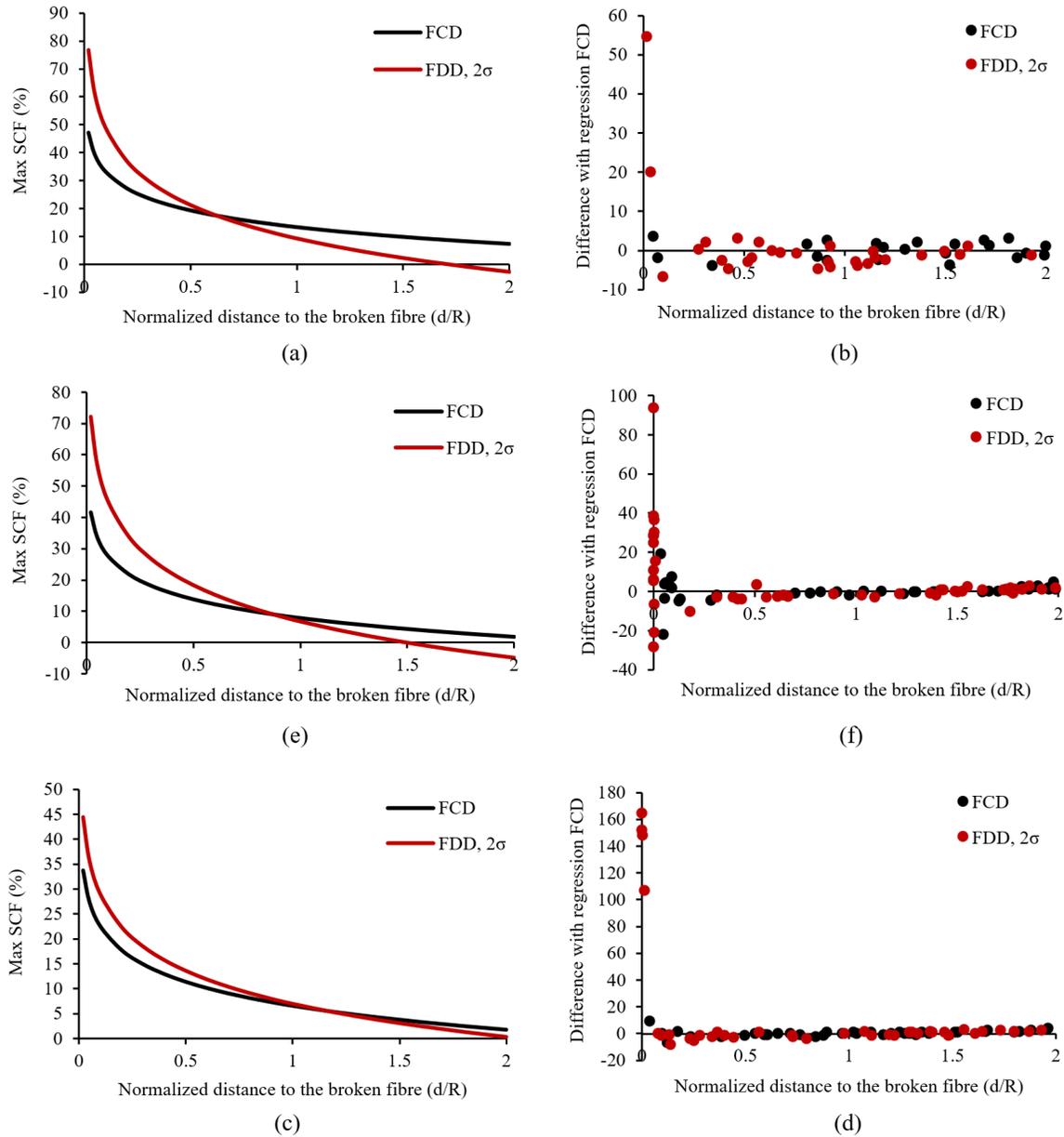

**Fig.9.** Stress concentrations based on normal peak stress for the FDD carbon T700s compared with the FCD bundles for different VFs at applied strain 2%, five random fibre placement realisations for each VF, each data point represents SCF in one intact fibre in one of the realisations of the FE model.

### 4.2.2. Average SCF vs Maximum SCF

The present section compares the two definitions of SCFs: the average SCF ($maxSCF_{avg}$) and the maximum SCF ($maxSCF_{max}$), computed over the fibre cross-section. The analysis is performed for both FCD and FDD bundles at applied strains of 0.1% (elastic regime) and 2% (inelastic regime), across different VFs.

Key observations are noted below:

1. **Magnitude of $maxSCF_{max}$ vs $maxSCF_{avg}$:** Fig. 10a–d illustrates that $maxSCF_{max}$ significantly exceeds $maxSCF_{avg}$ for both FCD and FDD bundles, regardless of the applied strain or VF. This disparity arises because $maxSCF_{max}$ captures the peak axial stress on the fibre surface, which is highly localized and can be several times greater than the average stress across the fibre cross-section (see section 5.1). In contrast, the actual stress concentrations that initiate fibre breaks are underestimated as $maxSCF_{avg}$ average out localized stress peaks. For instance, in Fig. 10a the $maxSCF_{max}$ in FCD bundles is approximately eight times greater than $maxSCF_{avg}$, emphasizing the critical role of stress localization.

2. **Elastic vs Inelastic Regime:** Two important trends can be seen when comparing elastic and inelastic deformation regimes:

   - **Elastic Regime (0.1% strain):** In FCD bundles, the peak $maxSCF_{max}$ is significantly higher (up to eight times) than in the inelastic regime (Fig.10a vs 10b). This is due to the continued matrix's elasticity which allows for efficient stress transfer and higher localized stresses near the break. For FDD bundles, the difference between elastic and inelastic regimes is less pronounced (approximately 25%, Fig. 10c vs 10d), as the variability in fibre diameters already introduces stress redistribution mechanisms that mitigate localized stress peaks.

   - **Inelastic Regime (2% strain):** Because of matrix yielding at applied strain 2%, the ability of the matrix to transfer shear stresses decreases, which reduces the matrix's capacity to transport shear stresses. This results in lower peak $maxSCF_{max}$ values compared to the elastic regime. However, the stress concentration decay is slower in the inelastic regime, as the matrix plasticity allows for a more distributed stress redistribution over a longer IL.

The significant difference between $maxSCF_{max}$ and $maxSCF_{avg}$ highlights the drawbacks of depending exclusively on average stress-based criteria in fibre break models. Although $maxSCF_{avg}$ offers a useful measure of overall stress redistribution, it is unable to identify the localized stress peaks that trigger the onset of fibre breaks. This is more important in FDD bundles, where diameter variability introduces additional stress localization to the fibre bundle. The results suggest that $maxSCF_{max}$ should be prioritized over the baseline models (e.g., [1–3]), as it better reflects the true stress concentrations that lead to fibre break.

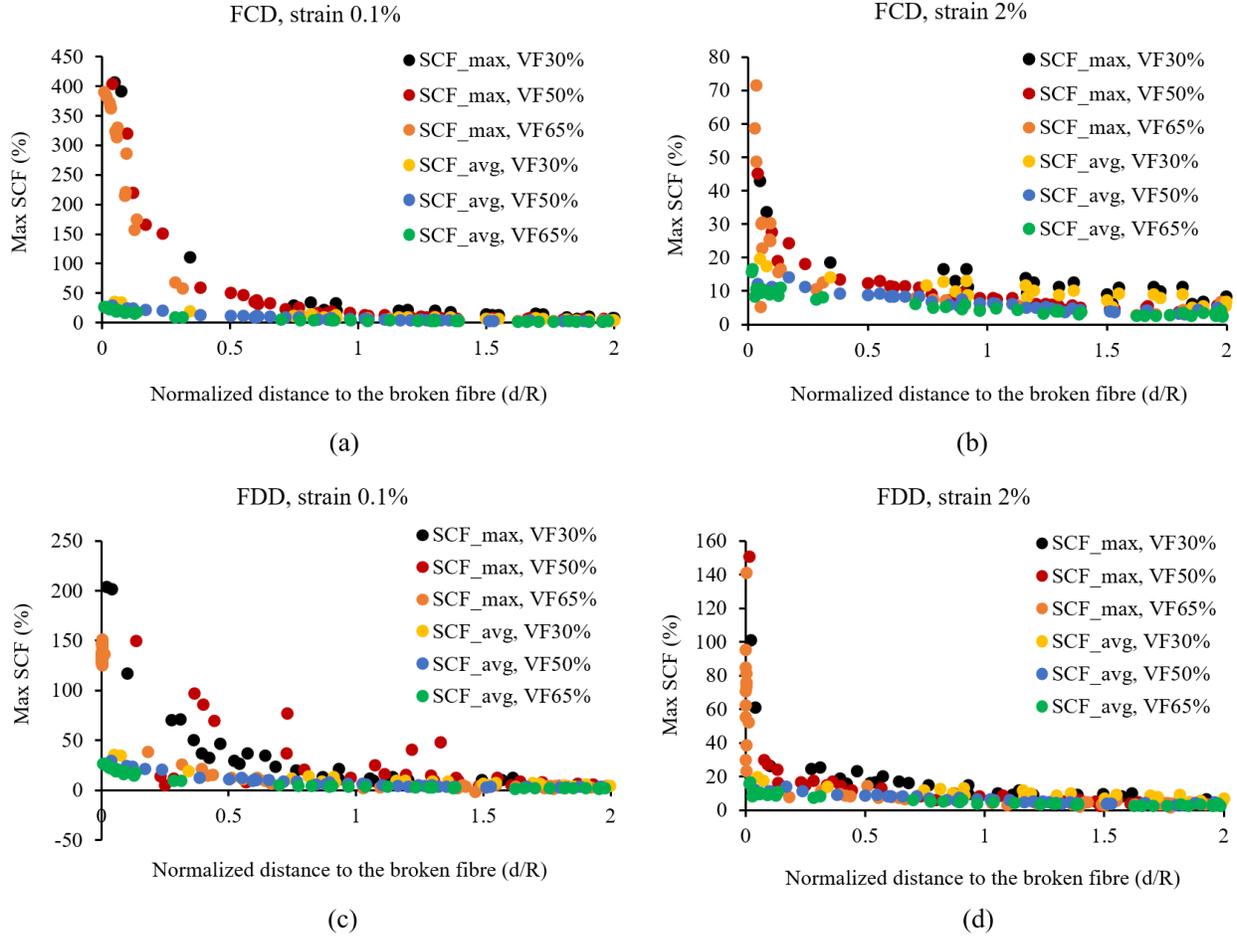

**Fig.10.** The maximum average SCF ($maxSCF_{avg}$) vs the peak stress in the cross-section ($maxSCF_{max}$): a, b) FCD bundle at applied strain 0.1% and 2%, respectively, and c, d) FDD bundle at applied strain 0.1% and 2%, respectively. Five random fibre placement realisations for each VF, each data point represents SCF in one intact fibre in one of the realisations of the FE model.

## 5. Discussion

### 5.1. Local SCF inside fibres

This section presents the local stress distribution over the cross-sections of four nearest intact fibres (Fibre-1 to Fibre-4) in the vicinity of a broken fibre. The study is performed for a FDD bundle with VF65% and a standard deviation of 2σ, at an applied strain of 2%.

Three important observations can be noted from Fig. 11: (I) **Stress Distribution Across Fibre Cross-Sections:** The findings represent that the closest fibre possesses greater (up to four times) axial stress than the nominal applied stress, which is reached further away from the surface. Again, this emphasizes the significance of calculating SCFs based on the maximum stress over the fibre cross-section ($maxSCF_{max}$) rather than the average stress ($maxSCF_{avg}$). Moreover, the values of $maxSCF_{max}$ are very strong at close distances to the broken fibre with a fast decay of stress concentration by increasing *d/R* (see also Fig.4b, 9, and 10). (II) **Influence of Fibre Diameter and Local Fibre Volume Fraction:** Fibre-1 vs Fibre-2: Although Fibre-1 is closer to the fiber break, its $maxSCF_{avg}$ is lower than Fibre-2. The denser local VF surrounding Fibre-1 explains this unexpected result, which enables nearby fibers to absorb a portion of the increased stress concentration. Fibre-3 vs Fibre-4: Despite the local VFs of both fibers are comparable,

Fibre-4 has a higher $maxSCF_{avg}$ than Fibre-3 due to its proximity to the fibre break. These illustrations exhibit how SCFs are determined by the interaction of local fiber packing and fiber proximity. **(III) Role of Fibre Diameter in Stress Localization:** When comparing Fibre-1 to Fibre-2 and Fibre-3 to Fibre-4, it is seen that the $maxSCF_{max}$ is higher in the fibers with a bigger diameter (Fibre-1 and Fibre-4), despite the latter having a smaller relative distance (*d/R*). This implies that fibres with bigger diameters are more prone to localized stresses due to their greater cross-section area, making them more susceptible to stress concentrations near breaks. On the other hand, fibres with smaller diameters experience lower SCFs and are overshadowed by bigger fibres that predominate close to the break plane.

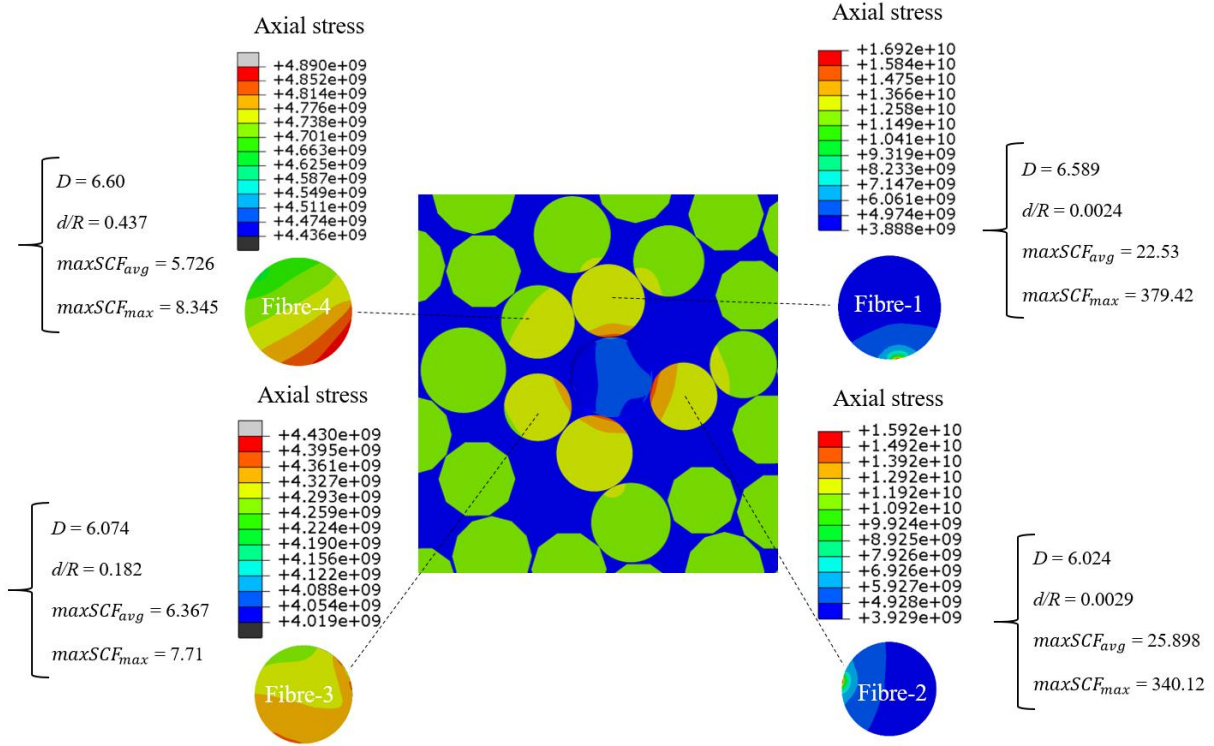

**Fig. 11.** Local stresses in four nearest intact fibres close to the broken fibre. The calculation was performed for a random realisation of FDD bundle with VF65% and standard deviation σ=2 at applied strain 2%. *D* denotes fibre diameter and *d/R* shows the relative distance from the broken fibre.

### 5.2. The significance of FDD on stress redistribution and cluster formation

In section SUPP 2 of the Supplementary materials, a statistical analysis t-Test is presented in order to evaluate the significance of the difference between FCD and FDD bundles spanning the entirety of the trend lines (*d/R*) based on average calculations. It supports the statements on the significance or insignificance of differences in SCF and IL, which can be made analysing the curves presented above for $maxSCF_{avg}$.

The $maxSCF_{avg}$ in NNFs for FDD with $2\sigma$ produce a reduction of –30% at VF65% and an increase of 13% and 18% at VF30% and 50% compared to FCD bundles, respectively. The $maxSCF_{max}$ in NNFs for FDD bundles demonstrated an increase of about 43-75% compared to FCD bundles (*d/R* < 0.5). The difference in few percent of the SCF value may play its role in appearing of new fibre breaks or break cascades.

As observed, the change in SCF for FDD bundles is much stronger for fibres which are the closest to the break (*d/R* < 0.5). This reflects a general fact that the microscopic stress distribution is highly sensitive to the fibre feature distribution (spatial, diameter, alignment) while the macroscopic elastic effective response of a laminate may not be affected by it [20]. Presence of small fibre diameter can lead to tighter packing [21]; there exists a negative correlation of the fibre modulus with the fibre diameter [19]. Presence of fibres with different diameters may affect local stress redistribution near a fibre break and hence cluster formation, even if the averaged SCF regression does not feel the difference.

Fig. 12 illustrates calculation of $maxSCF_{avg}$ in the closest neighbours of a broken fibre for five random realisations of carbon bundles with $\sigma$ at VF50%, which all have different diameters. The mean diameter ($\bar{\mu}$) of carbon fibres is considered to be 6.76 μm. It shows carbon bundle with 18 fibre diameters smaller than $\bar{\mu}$, 21 fibre diameters bigger than $\bar{\mu}$, and 2 fibre diameters equal to $\bar{\mu}$ value. The calculated SCF mean difference produces a positive value (Table 5, bold values). This suggests that when a fibre break occurs, the distribution of SCF in neighbouring fibres is affected due to the accumulation of either smaller or bigger fibre diameters. The increased SCF in NNFs arises from an accumulation of bigger fibre diameters, and vice versa (see section 5.1 as well).

Table 5, for more demonstration, presents the total number of fibres (BF plus NNFs) in five random realisations in all FDD bundles ($\sigma/2$, $\sigma$, $2\sigma$), binned by their diameters. It can be noticed that the SCF mean differences exhibit negative and positive values for carbon/epoxy bundles. As it can be observed, the SCF mean difference is positive where there are more fibre diameters bigger than μ, but smaller values result in negative differences.

**Table. 5.** Diameter distribution of fibres (BF and NNFs together) in FDD carbon/epoxy bundle realisations, as well as corresponding SCF difference mean values from statistical analysis as presented in SUPP 2.

| Fibre | VF | Number of total (BF plus NNFs) fibres in five random realisations, binned by their diameters | | | | | | | | | SCF relative difference $\bar{\mu}$ ($\sigma/2$, $\sigma$, $2\sigma$ – RFCD)/RFCD % | | |
|---|---|---|---|---|---|---|---|---|---|---|---|---|---|
| | | $< \bar{\mu}$ | | | $> \bar{\mu}$ | | | $= \bar{\mu}$ | | | FCD vs $\sigma/2$ | FCD vs $\sigma$ | FCD vs $2\sigma$ |
| | | $\sigma/2$ | $\sigma$ | $2\sigma$ | $\sigma/2$ | $\sigma$ | $2\sigma$ | $\sigma/2$ | $\sigma$ | $2\sigma$ | | | |
| Carbon | 30% | 9 | 20 | 11 | 24 | 16 | 20 | 5 | 3 | 4 | 0.6 | -2.4 | 0.2 |
| | 50% | 23 | **18** | 5 | 12 | **21** | 26 | 6 | **2** | 3 | -4.2 | **0.6** | 0.5 |
| | 65% | 21 | 33 | 25 | 9 | 6 | 0 | 3 | 4 | 2 | -6.2 | -6.4 | -3.5 |

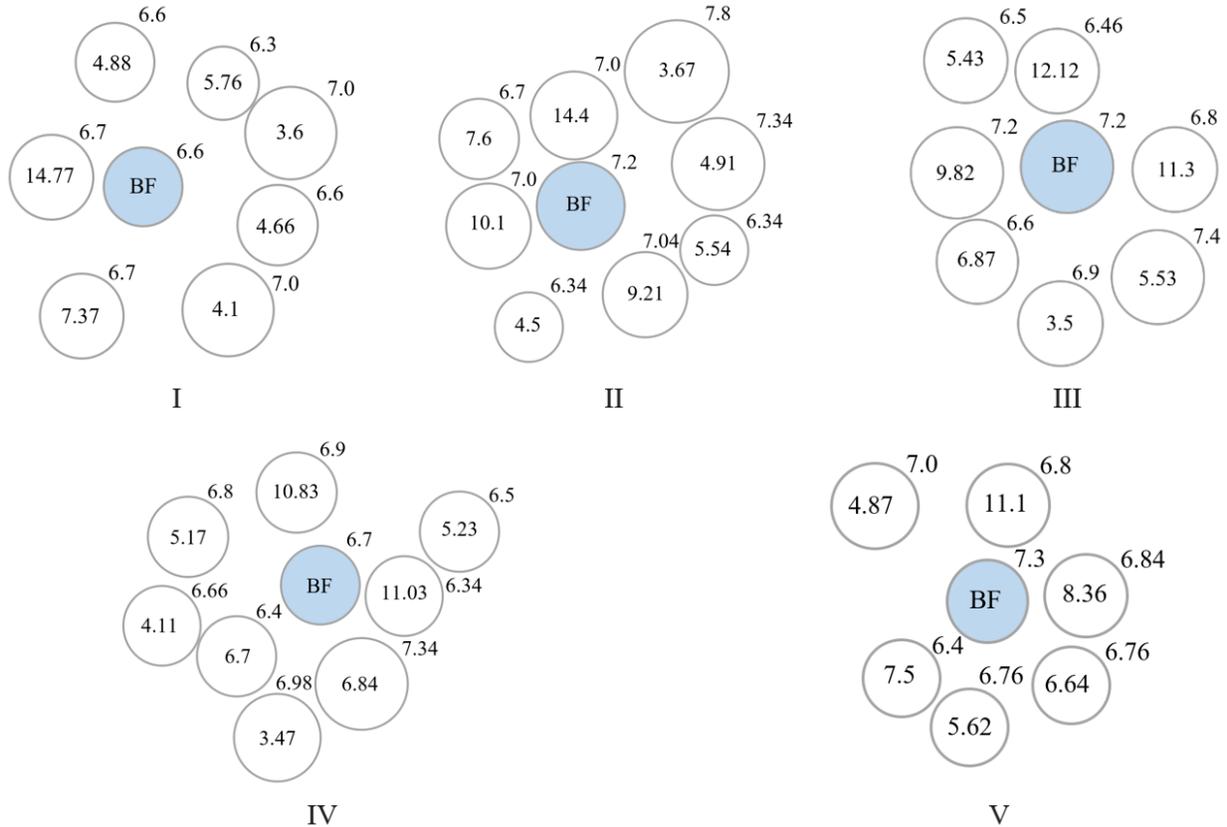

**Fig.12.** Five FDD random realisations of carbon fibre bundle with $\sigma$ at VF50%. Blue circle shows the broken fibre (BF) and grey ones depict the nearest neighbour fibres (NNF). The values outside (top-right) and inside the circles correspond to the fibre diameter in μm and the SCF derived in FE models, respectively.

## 6. Conclusion

This study evaluates the effect of fiber diameter variability on stress redistribution (SCF and IL) following a single fiber break in a bundle of unidirectional (UD) impregnated fibers. A finite element analysis of the stress condition of a bundle surrounding a broken fiber served as the basis for the study. Carbon fibre/epoxy bundles with experimentally measured fibre diameter distributions (FDD) were used in the analysis, followed by a parametric study to determine the influence of the FDD coefficient of variation on SCF and IL. For each fibre volume fraction value of VF30%, VF50%, and VF65%, five random realisations were built. SCFs, ILs and stresses in NNFs were assessed: (I) at strain levels of 2% and 0.1% (inelastic and elastic regime), and (II) average and maximum normal stress over the cross-section.

Key findings reveal that FDD alters stress redistribution around fibre breaks, with critical implications for failure prediction:

1. **SCF Dependency on Fibre Diameter and Clustering:** Bigger fibre diameters increase stress concentration factor (SCF) in nearest neighbour fibres (NNFs), while clusters of such fibres intensify stress concentrations. In contrast, fibres with smaller diameters mitigate SCF, highlighting the role of local microstructure in failure initiation.

2. **Critical Role of Maximum Stress Criteria:** Peak stress criteria should be prioritized in bundle strength models to accurately capture failure risks because SCF computed using maximum cross-sectional stress ($maxSCF_{max}$) significantly exceed average stress-based values ($maxSCF_{avg}$), with differences ranging from 40% to 75% in NNFs for FDD bundles compared with FCD bundles. This challenges conventional benchmarks relying on averaged SCF values.

3. **Deformation Regime Effects:** Matrix plasticity (2% applied strain) diminishes stress concentrations compared to the elastic regime while extending ineffective lengths about twice more than the elastic regime.

To our knowledge, this is the first study to quantify the impact of FDD variability on localized stress fields, bridging a gap in micromechanical models. Implementing the FDD-calculated SCF based on maximum stress over the fibre's cross-section is important for improving a bundle strength model, the failure strain of UD composites can be more realistically predicted.